\documentclass[aps,epsf,twocolumn,prl,showpacs]{revtex4}
\usepackage{amsmath}
\usepackage{amsfonts}
\usepackage{amssymb}
\usepackage{graphicx}
\usepackage[scanall]{psfrag}

\begin{document}
\title{The Anomalous Behavior of Solid $^{4}$He in Porous Vycor Glass}

\author{Xiao Mi}

\author{John D. Reppy}
\email{jdr13@cornell.edu}

\affiliation{Laboratory of Atomic and Solid State Physics and the Cornell Center for Materials Research, Cornell University, Ithaca, New York 14853-2501}

\begin{abstract}
The low temperature properties of solid $^4$He contained in porous Vycor glass have been investigated utilizing a two-mode compound torsional oscillator. At low temperatures, we find period shift signals for the solid similar to those reported by Kim and Chan \cite{ref1}, which were taken at the time as evidence for a supersolid helium phase. The supersolid is expected to have properties analogous to those of a conventional superfluid, where the superfluid behavior is independent of frequency and the ratio of the superfluid signals observed at two different mode periods will depend only on the ratio of the sensitivities of the mode periods to mass-loading. In the case of helium studies in Vycor, one can compare the period shift signals seen for a conventional superfluid film with signals obtained for a supersolid within the same Vycor sample. We find, contrary to our own expectations, that the signals observed for the solid display a marked period dependence not seen in the case of the superfluid film. This surprising result suggests that the low temperature response of solid $^4$He in a Vycor is more complex than previously assumed and cannot be thought of as a simple superfluid.
\end{abstract}

\date{April 23, 2012}

\maketitle
\narrowtext

\begin{figure}
\includegraphics[width=1\columnwidth]{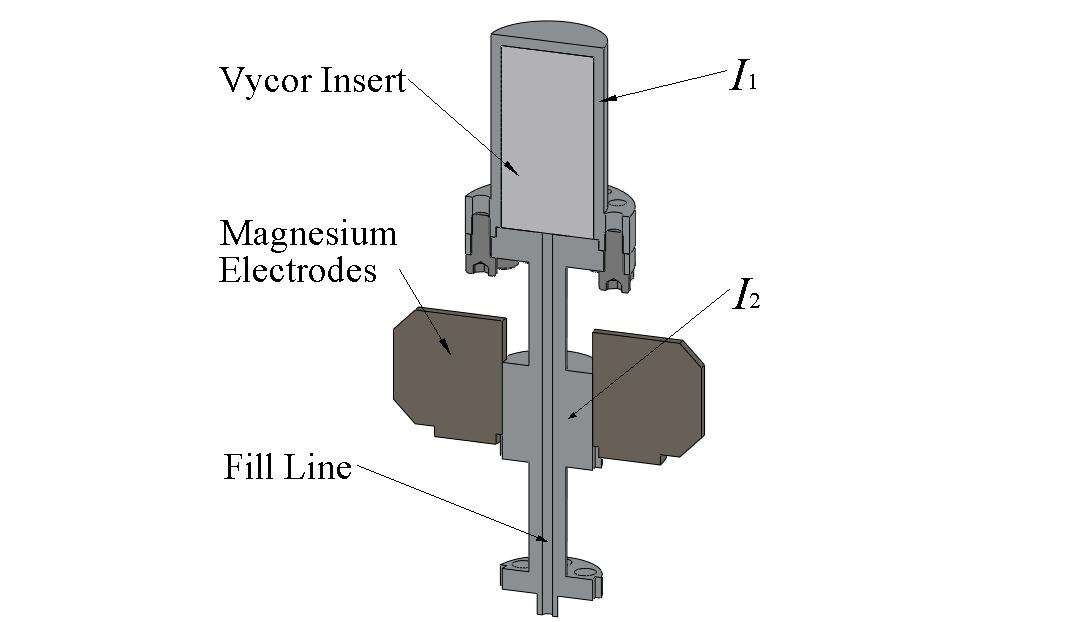}
\caption{Cross section of the compound vycor oscillator}\label{fig1}
\end{figure}

In 2004, Kim and Chan (KC) \cite{ref1}, employing a torsional oscillator (TO) containing a sample of porous Vycor glass impregnated with solid $^4$He, reported the first clear evidence for a possible $^4$He supersolid phase. In their measurement they found an anomalous decrease in the period of the oscillator as the temperature was lowered below about 200 mK. This period-shift signal was interpreted as evidence for a superfluid-like decoupling of a fraction of the moment of inertia of the solid $^4$He from the TO. Such a mass decoupling or non-classical rotational inertia (NCRI) signal is a characteristic of the supersolid state \cite{ref2}. At the same time, they also reported a sensitivity of the period-shift signals to the level of $^3$He impurities and a “critical velocity” effect where the signal size was substantially reduced by increases in the drive level. In a second publication \cite{ref3}, KC extended their measurements to bulk solid and found an essentially identical supersolid phenomenology to that observed in solid $^4$He in Vycor including the sensitivity to drive amplitude and $^3$He impurity level. Given the four orders of magnitude difference in the confining dimension of the Vycor pores as compared to the millimeter scale geometry for the bulk experiments, this similarity between the Vycor and bulk $^4$He results is remarkable.

A second important development was the observation by Day and Beamish (DB) \cite{ref4} of an anomalous increase in the elastic shear modulus of the solid $^4$He extending over the same temperature range as the supersolid phenomenon and also accompanied by a sensitivity to $^3$He impurity level and dependence on the drive level similar to that seen in the KC supersolid experiments. This similarity suggests a close association between the elastic anomaly and the supersolid phenomenon. In the case of bulk samples in relatively large containers, the acceleration of the solid during oscillation through the elastic interaction between the solid and the TO might lead to observable effects. However, in the case of Vycor it is difficult to believe that the shear modulus could play a significant role given the rigidity of the Vycor glass and the small dimensions of pores confining the solid $^4$He.

A recent torsional oscillator (TO) study \cite{ref5} of bulk samples of solid $^4$He utilizing a multiple frequency oscillator showed that period shift signals observed with this oscillator, although similar in form to classic supersolid period shift signals, are in fact attributable in large part to the temperature dependence of the shear modulus. Measurements at two different oscillator periods allow a separation of the observed period shift signals into two components: first a period-independent contribution, such as that expected for the supersolid or a conventional superfluid, and second a period-dependent contribution due to dynamic effects.  The period shift signal arising from the shear modulus anomaly is expected to be negligible for solid $^4$He in Vycor, so a compound oscillator study of this system should provide definitive test of the nature of the signals observed by KC. The compound oscillator we have constructed for the current series of Vycor measurements is shown in Figure~\ref{fig1}. Our oscillator design is similar to that of Aoki, Graves and Kojima \cite{ref6} who pioneered the use of the compound oscillator in supersolid studies of the bulk solid.

The oscillator is largely constructed from high strength aluminum alloy and consists of two moments of inertia and two nearly equal torsion rods, 0.5715 cm in outer diameter and 0.127 cm in inner diameter. The entire oscillator structure is rigidly attached to a massive block of Cu which is thermally anchored to a dilution refrigerator. The torsion constant for the upper torsion rod is $k_1 = 1.41 \times 10^9$ dyncm and $ k_2 = 1.65 \times 10^9$ dyncm for the lower rod. The upper moment of inertia, $ I_1 = 20.95$ gcm$^2$, contains a sample of porous Vycor glass sealed with epoxy in an aluminum cylinder with an outer diameter of 1.778 cm. The Vycor sample is slightly elliptical with a mean diameter of 1.43 cm and length 2.96 cm. The lower moment of inertia, including the magnesium electrodes used for exciting and detecting the two rotational modes of the oscillator, is $I_2 = 14.55$ gcm$^2$. The oscillator structure has two rotational modes, an upper mode (designated by $+$) where the moments of inertia execute a counter rotational (anti-phase) motion and a lower mode (designated by $-$) where the two moments of inertia rotate together in-phase. The low temperature oscillator periods are $P_+ = 0.389$ ms and $P_- = 1.112$ ms for the high and low modes, respectively. The periods of oscillation for the modes are given by $P_\pm = 2\pi\{[\frac{k_1(I_1 + I_2) + k_2I_1}{2I_1I_2}][1 \pm \sqrt{1 - \frac{4k_1k_2I_1I_2}{(k_1(I_1 + I_2) + k_2I_1)^2}}]\}^{-\frac{1}{2}}$.

The electrodes mounted on the oscillator are connected to a low-impedance 220 V source. Since we employ a current amplifier to detect the motion of the electrodes, the recorded signal amplitude is proportional to the angular velocity, $\dot{\theta}_2$. The relation between the angular velocity, $\dot{\theta}_1$, of the Vycor sample and that of the electrodes, obtained from the compound oscillator equations of motion is $\dot{\theta}_1 = [(\frac{k_2}{I_1})(\frac{P_\pm}{2\pi})-\frac{I_2}{I_1}] \dot{\theta}_2$. For the high ($+$) mode, $\dot{\theta}_1 = -0.381\dot{\theta}_2$, and for the low ($-$) mode, $\dot{\theta}_1 = 1.848\dot{\theta}_2$. Thus the angular velocity of the sample is a factor of 4.5 larger for the low mode as compared to the high mode for a given detected signal amplitude.

Commercial $^4$He gas, with a nominal 0.3 ppm $^3$He impurity level, is used for the samples that are formed by the blocked-capillary method. Typically, the cell is loaded at a pressure of 70 bar ($\sim$1000 psi) at a temperature of 3 K. Upon cooling, the fill line to the cell freezes before the helium in the Vycor. After the freezing of the $^4$He within the Vycor, the sample pressure is approximately 40 bar.

It is our usual practice to excite both modes simultaneously and employ separate lock-in amplifiers to record the individual response for each mode. A feed-back system is used to control the drive voltage and maintain the signal amplitudes at a constant level, thus minimizing any non-linear amplitude-dependent effect. The two modes have different sensitivities for changes in the moment of inertia of the $^4$He sample. These sensitivities were experimentally determined from the period shift data, $\delta P_\pm$, observed while condensing measured volumes of $^4$He in the Vycor sample. The shifts, following filling of the Vycor with solid, are $\delta P_+=156.4$ ns for the high mode and $\delta P_-=1426.6$ ns for the low mode. The contribution of the solid $^4$He, $I_{\text{He}}$, to the torsion bob moment of inertia, $I_1$, can also be obtained from the filling data and is found to be approximately $8.1 \times 10^{-2}$ gcm$^2$. The mass-loading sensitivities are $m_+ = 1.931 \times 10^3$ ns/(gcm$^2$) for the high mode and $m_- = 17.612 \times 10^3$ ns/(gcm$^2$) for the low mode. The sensitivity ratio is $m_+/m_-=$ 0.11.

\begin{figure}
\includegraphics[width=1\columnwidth]{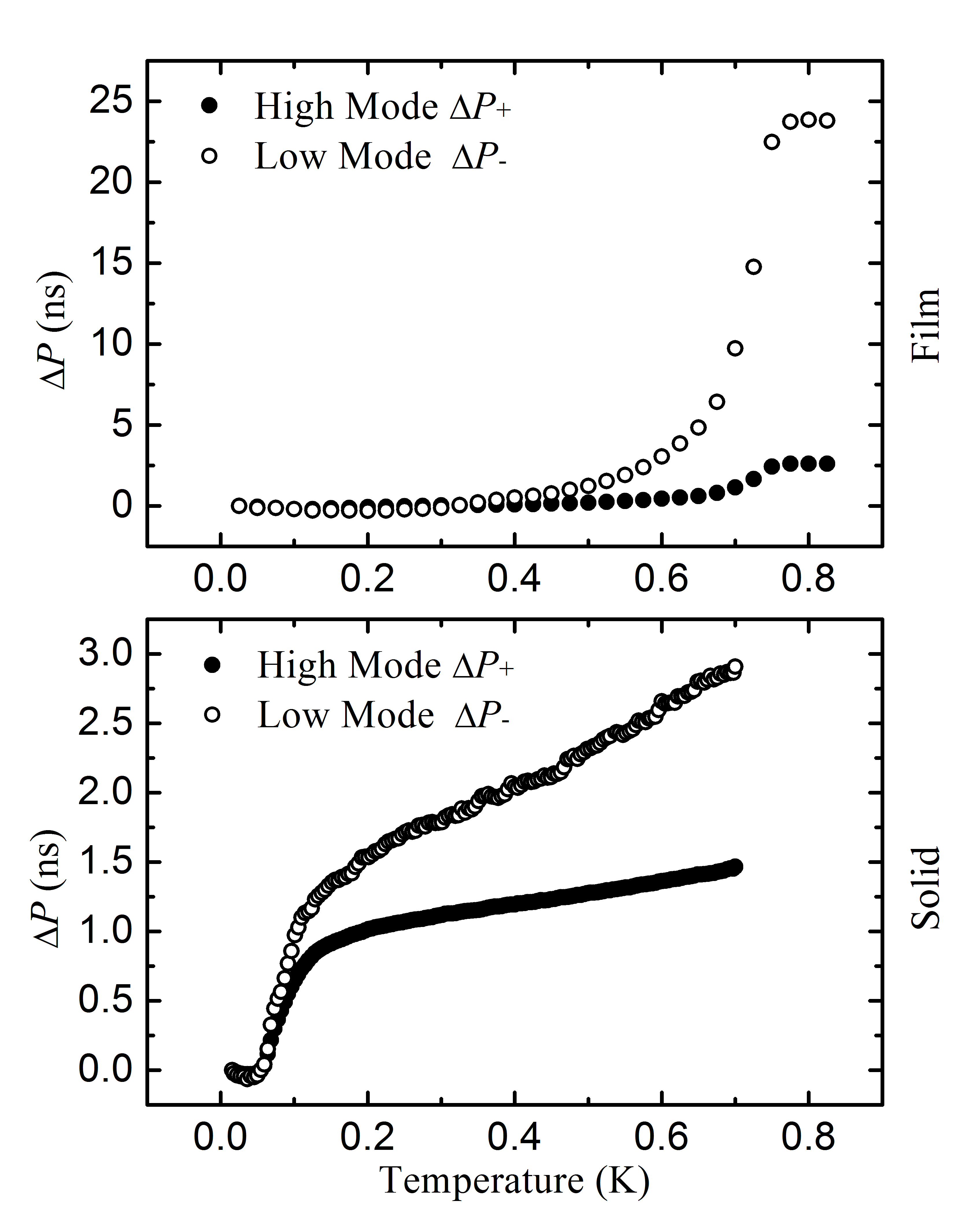}
\caption{The upper panel shows the period shift data for a superfluid film adsorbed in the Vycor sample. The lower panel shows comparable data for the case of $^4$He frozen within the Vycor. Note the difference in the vertical axis scales.}\label{fig2}
\end{figure}

For the superfluid film data shown in Figure~\ref{fig2}, the period difference between the mode period at the transition temperature of 750 mK and the period data at lower temperatures, $\Delta P_{\text{Film}} (T) = P_{\text{Film}} (750 \text{ mK}) - P_{\text{Film}} (T)$, is proportional to the superfluid mass of the adsorbed film. The striking aspect of the film data is the large ratio between the magnitudes of the high and low mode signals. However, this is just what should be expected for a superfluid on the basis of the mass loading calibration. The lower panel shows comparable data for the case of solid $^4$He within the Vycor sample. Here the situation is quite different and there is only factor of approximately 2 between the magnitudes of the signals for the two modes. Both the superfluid film data and the solid data have been corrected for the empty cell temperature-dependent backgrounds.

Although both modes for the solid show the same anomalous drop in the oscillator period as that reported by KC for solid $^4$He in Vycor, the ratio of the magnitudes of the signals for the two modes is much less than what would be expected from the mass loading calibration. To emphasize this point we treat the period change, $\Delta P_{\text{Sol}} = P_{\text{Sol}} (150 \text{ mK}) - P_{\text{Sol}}(20 \text{ mK})$, between 150 mK and 20 mK as a standard NCRI signal. These changes in period are 1.354 ns and 0.916 ns for the low and high mode respectively.  We can calculate a supersolid fraction or NCRIF in the usual way by dividing these period shifts by the corresponding period shift values obtained when filling the cell with solid. The NCRIF value for the low mode is then 0.095 \%, while for the high mode value it is 0.586 \%. For a genuine superfluid signal these values should be equal and independent of the TO frequency. Thus we are forced to consider other mechanisms for the observed period shifts for the solid beyond the usual supersolid scenario.

\begin{figure}
\includegraphics[width=1\columnwidth]{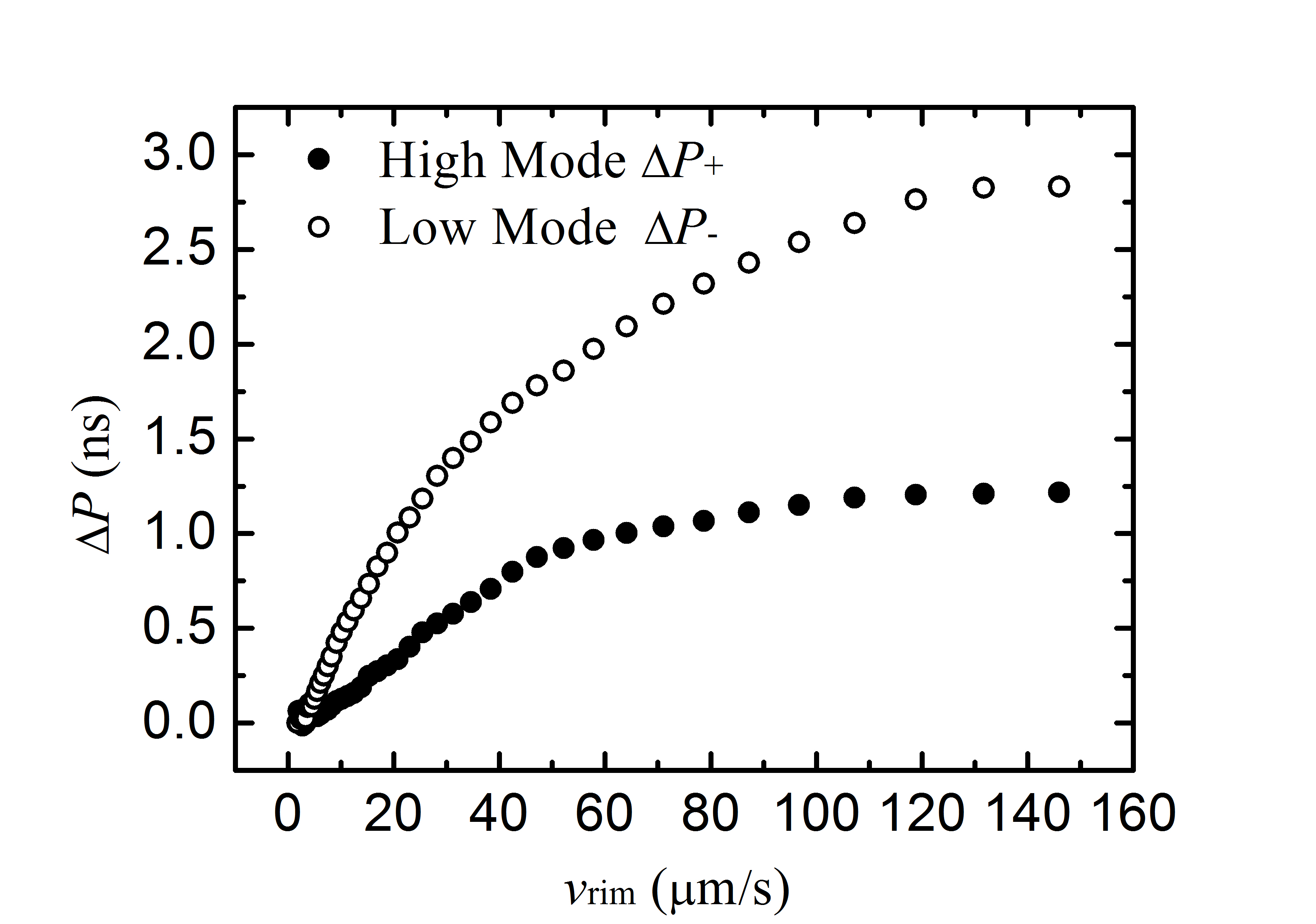}
\caption{The increase with velocity in the mode periods at the temperature of 20 mK are plotted as a function of the maximum rim velocity.}\label{fig3}
\end{figure}

In addition to the temperature dependence of the solid signals, we have also investigated their sensitivities to increasing velocity. We find, in agreement with KC, that as the angular velocity of the torsion bob is increased, the low temperature values of the mode periods increase, thus decreasing the overall magnitude of the solid period shift signals. In Figure~\ref{fig3} we plot the values, obtained at 20 mK, for the increase in mode periods with increasing total angular velocity as functions of the maximum rim velocity. This data set was obtained while driving both modes of the oscillator. In contrast to what is often seen in bulk $^4$He measurements of this type, there is no evidence in these measurements of a subcritical velocity region where the period shift signal is independent of the rim velocity. In our data, the initial change in period signal appears to be linearly proportional to the velocity starting from our lowest velocity.

\begin{figure}
\includegraphics[width=1\columnwidth]{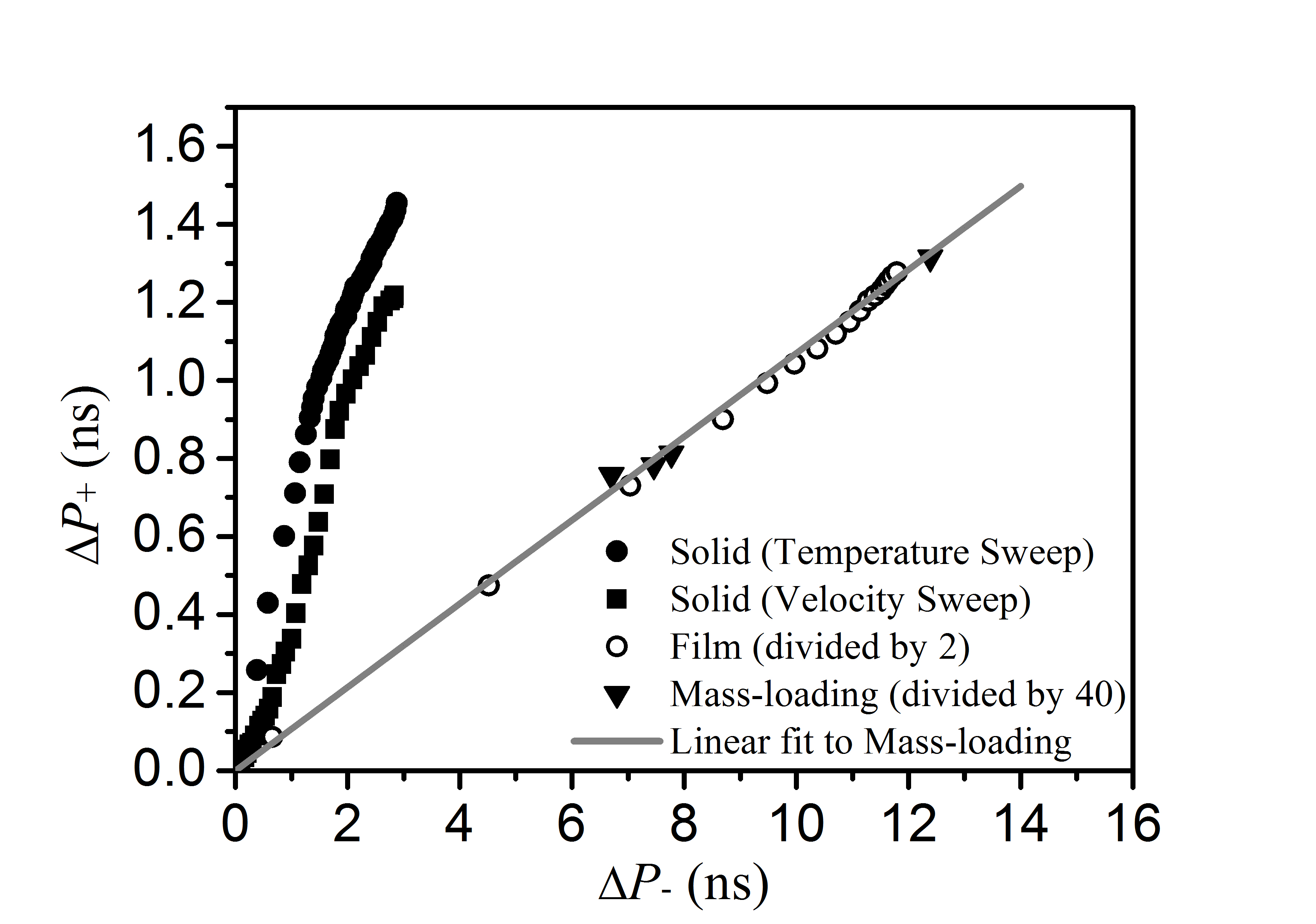}
\caption{In this figure the period shift data for the two modes displayed in Figures~\ref{fig2} and~\ref{fig3} are plotted against each other. In addition, data obtained while incrementally filling the cell are also shown. The period shift values for the film and mass loading data sets have been divided by the factors of 2 and 40 respectively. }\label{fig4}
\end{figure}

An interesting way to display these data as well as the period shift data for the temperature dependent solid signals is to plot the period shifts for the two modes against each other. This representation of the data is shown in Figure~\ref{fig4}. Also included in this figure are film data from the transition temperature down to 300 mK and period shift data obtained while filling the cell with $^4$He. In contrast to the superfluid film data, which agree quite well with the mass-loading calibration as indicated by the straight line in the figure, the solid data follow a much steeper trend. The velocity and temperature sweep data show a very similar behavior. This similarity gives support to the hypothesis \cite{ref7} of the Davis group at Cornell that velocity and temperature play complementary roles in controlling the low temperature dynamics of solid $^4$He. The solid signals show an unanticipated sensitivity to the frequency of the measurement and bear a resemblance to the period shift signals due to elastic modulus effects seen in earlier bulk solid experiments \cite{ref5}.

There are two obvious ways in which the temperature dependent shear modulus might influence this experiment. The first would be the influence of the temperature dependent shear modulus on the dynamic response of the torsion bob itself, and second, the effect due to the temperature dependence of the shear modulus of the solid $^4$He in the torsion rods of the compound oscillator. An explanation based on possible shear modulus changes of the solid within the Vycor seems unlikely, however, given the rather large value of the Vycor shear modulus, on the order of $7 \times 10^9$ Pa, as compared to the value of $1.5 \times 10^7$ Pa for the solid $^4$He. It is known, however, from acoustic measurements \cite{ref8,ref9} that the elastic modulus of a Vycor sample containing $^4$He increases by about 0.3 \% upon freezing of $^4$He. An estimate based on a shear modulus change of this magnitude leads to period shifts of less than 0.1 ns, much smaller than the observed signals. In the case of period shifts due to the temperature dependent changes of the shear modulus of the solid within the torsion rods, the period shifts are somewhat larger when assuming a 100 \% change in the shear modulus, on the order of 0.19 ns for the high mode and 0.55 ns for the low mode. These period shifts are larger than the shifts due to shear modulus effect on the dynamics of the TO, but are still relatively small compared with the observed period shift signals for the solid. When corrected for the mass loading sensitivity factors, these period shifts yield a frequency independent contribution to the effective moment of inertia, $\Delta I_{\text{eff}} = 1.035 \times 10^{-5}$ gcm$^2$, for an equivalent NCRIF of about $1.26 \times 10^{-4}$.

\begin{figure}
\includegraphics[width=1\columnwidth]{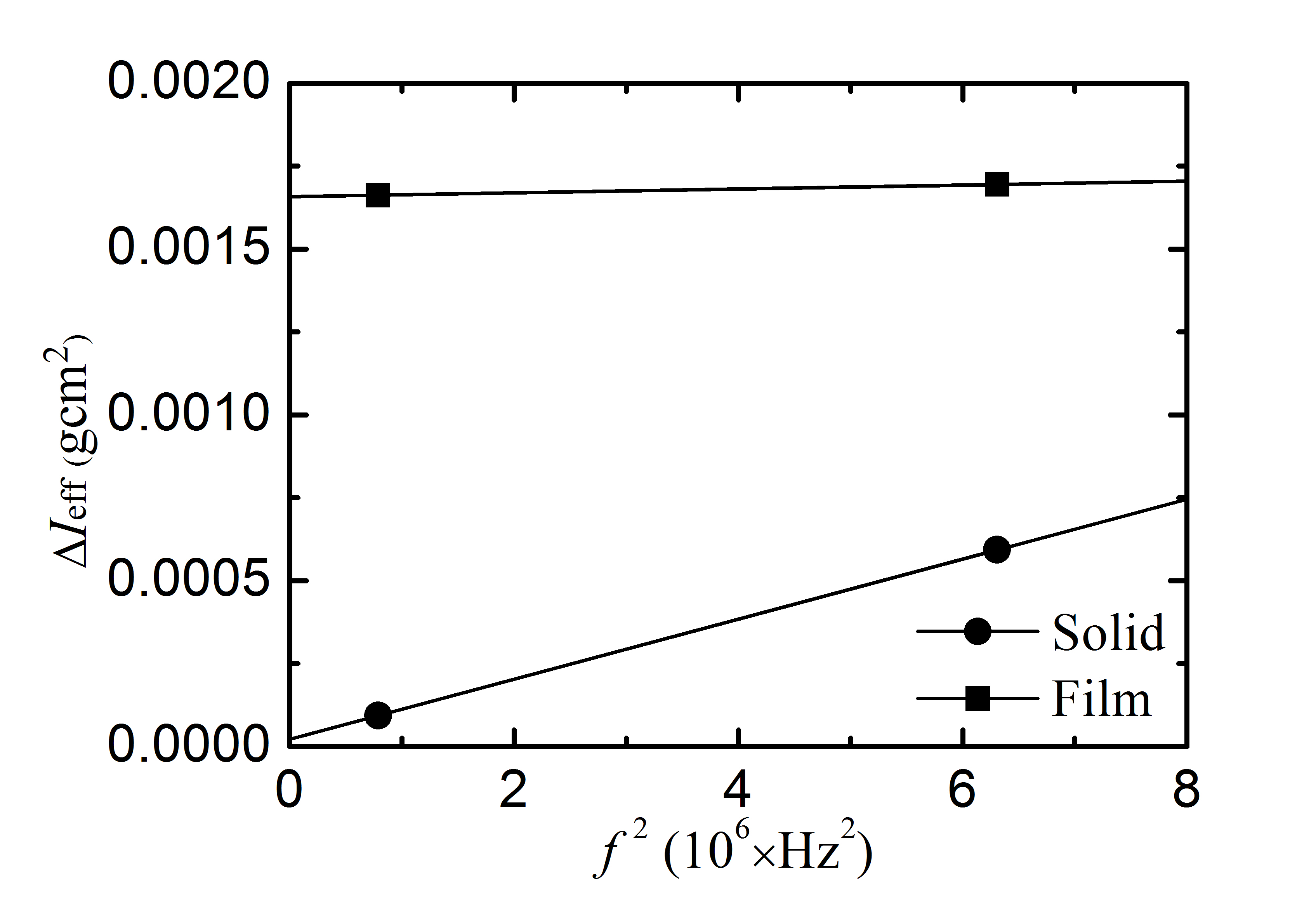}
\caption{This figure examines the relationship between the effective moment of inertia, $I_{\text{eff}}$, and the square of the mode frequency. The zero frequency intercept represents a superfluid-like (i.e. frequency independent) contribution to the total signal.}\label{fig5}
\end{figure}

Although our estimates of the possible period shifts due to the elastic properties of the $^4$He are too small to account for the observed period shifts, the strong frequency dependence of both the velocity and temperature sweep signals suggests that there may be some other, as yet unknown, dynamic process coming into play. In Figure~\ref{fig5} we analyze the data in terms of a simple model consisting of a dynamic contribution to the effective moment of inertia for each mode proportional to the frequency squared and a frequency independent term corresponding to a possible NCRI contribution. In this figure the effective moment of inertia values derived from the temperature dependent period shift signals at 150 mK are plotted against the square of the mode frequency. In this model the zero frequency intercept of the line, $\Delta I_{\text{eff}} (f^2) = (2.23 + 9.04f^2) \times 10^{-5}$ gcm$^2$, determined by our two data points representing the frequency independent or NCRI contribution to the moment of inertia. For the data shown the intercept is small but positive and would correspond to a NCRIF of about $2.72 \times 10^{-4} \pm 3.2 \times 10^{-4}$.  The error estimate was made assuming a resolution of  $\pm$0.05 ns in the period shift data. One would then conclude from these data that there is no evidence for a significant contribution from a supersolid term to the Vycor period shift data. This conclusion must be treated with caution since we do not know the origin of the frequency dependence of the data or even, as has been assumed in the foregoing analysis, that this dependence follows the square of the frequency. Clearly measurements at least three different frequencies are mandated.

In conclusion, we have examined the behavior of a Vycor sample containing solid $^4$He mounted on a two-frequency compound TO. Period shift signals similar to those reported by KC are observed. The response at the two different frequencies, however, demonstrates a pronounced dependence on frequency not expected for a true superfluid. The temperature dependent shear modulus of the solid does not appear to lead to large enough period shifts in the oscillator to explain the observations. Thus we are left with an as yet unexplained phenomenon for solid $^4$He in the porous Vycor glass. This problem will require further experimental and theoretical effort to achieve a solution.

\begin{acknowledgements}
Xiao Mi wishes to thank the Hunter R. Rawlings III Cornell Presidential Research Scholarship for its continuing support. The authors acknowledge useful and encouraging discussions with J.R. Beamish, M.H.W. Chan, H. Kojima and Erich J. Mueller. This work was supported by the National Science Foundation through Grant DMR-060586 and CCMR Grant DMR-0520404.
\end{acknowledgements}

\end{document}